\pdfoutput=1
\documentclass[a4paper]{article}

\usepackage{INTERSPEECH2020}
\usepackage{amsmath,graphicx}

\usepackage{hyperref}
\graphicspath{{grf/}}

\newcommand{\x}{\ensuremath{\mathbf{x}}}
\newcommand{\y}{\ensuremath{\mathbf{y}}}

\newcommand{\bbP}{\ensuremath{\mathbb{P}}}

\newcommand{\calB}{\ensuremath{\mathcal{B}}}

\newcommand{\calL}{\ensuremath{\mathcal{L}}}

{%
\begin{list}{#1}{
\vspace{-\topsep}
\vspace{-\partopsep}
\setlength{\itemindent}{0cm}
\setlength{\rightmargin}{0cm}
\setlength{\listparindent}{0cm}
\settowidth{\labelwidth}{#1}
\setlength{\leftmargin}{\labelwidth}
\addtolength{\leftmargin}{\labelsep}
\setlength{\itemsep}{0cm}
}%
}%
{%
\end{list}
\vspace{-\topsep}
\vspace{-\partopsep}
}

{\begin{enumerate}%
}%
{\end{enumerate}}

\usepackage{color}
\usepackage{colortbl}
\definecolor{light-gray}{gray}{0.8}

\title{Semi-supervised ASR by End-to-end Self-training}
\name{Yang Chen$^{\star}$ \thanks{Work done while Yang Chen was an intern at Amazon Alexa.} \hspace*{4em} Weiran Wang$^{\dagger}$\thanks{Work done while Weiran Wang was at Amazon Alexa.} \hspace*{4em} Chao Wang$^{\dagger\dagger}$}
\address{\hspace*{-1em} $^{\star}$Georgia Institute of Technology \hspace*{0.7em} $^{\dagger}$Salesforce Research \hspace*{0.7em} $^{\dagger\dagger}$Amazon Alexa}
\email{\hspace*{1em} ychen3411@gatech.edu \hspace*{0.3em} weiran.wang@salesforce.com \hspace*{0.3em} wngcha@amazon.com}

\begin{document}
\maketitle

\begin{abstract}
While deep learning based end-to-end automatic speech recognition (ASR) systems have greatly simplified modeling pipelines, they suffer from the data sparsity issue. 
In this work, we propose a self-training method with an end-to-end system for semi-supervised ASR. 
Starting from a Connectionist Temporal Classification (CTC) system trained on the supervised data, we iteratively generate pseudo-labels on a mini-batch of unsupervised utterances with the current model, and use the pseudo-labels to augment the supervised data for immediate model update.
Our method retains the simplicity of end-to-end ASR systems, and can be seen as performing alternating optimization over a well-defined learning objective. We also perform empirical investigations of our method, regarding the effect of data augmentation, decoding beamsize for pseudo-label generation, and freshness of pseudo-labels. 
On a commonly used semi-supervised ASR setting with the Wall Street Journal (WSJ) corpus, our method gives 14.4\% relative WER improvement over a carefully-trained base system with data augmentation, reducing the performance gap between the base system and the oracle system by 46\%.
\end{abstract}

\section{Introduction}
\label{s:intro}
\vspace*{-1ex}

One challenge faced by modern ASR systems is that, with ever enlarged model capacity,  large amount of labeled data are required to thoroughly train them. Unfortunately, collecting and transcribing huge dataset is expensive and time-consuming. As a result, semi-supervised ASR has been an important research direction, with the goal of leveraging a large amount of unlabeled data and a much smaller amount of labeled data for training. 
One of the simplest methods in this setting is self-training, which uses the decoding results or pseudo-labels on unsupervised data, often at the word level, to augment supervised training. It has been shown to be very effective with traditional ASR pipelines~\cite{long_asr,semi_conf, semi_conf2, amz_1m}. 

In this work, we propose a novel framework for self-training in an end-to-end fashion. Starting from a carefully-trained Connectionist Temporal Classification (CTC,~\cite{Graves_06a}) system, we alternate the following two procedures: generating pseudo-labels using a token-level decoder on a mini-batch of unsupervised utterances, and augmenting the just decoded (input, pseudo-label) pairs for supervised training.  We show that this method can be derived from alternating optimization of a unified objective, over the acoustic model and the non-observed labels of unsupervised data. The two procedures effectively reinforce each other, leading to increasingly accurate models. 

We emphasize a few important aspects of our method, which distinguish our work from others (detailed discussions on related work are provided later):
\begin{itemize}
	\item The pseudo-labels we use are discrete, token-level label sequences, rather than per-frame soft probabilities.
        \item The pseudo-labels are generated on the fly, rather than in one shot, since fresh labels are of higher quality than those produced from a stale model.
        \item We perform data augmentation not only on supervised data, but also on unsupervised data.
\end{itemize}
These modeling choices, which lead to performance gain over alterantives, are backed up by empirical results. 
We demonstrate our method on the WSJ corpus (\cite{PaulBaker92a}, LDC catalog numbers LDC93S6B and LDC94S13B). 
Our method improves PER by 31.6\% relative on the development set, and WER by 14.4\% relative on the test set from a well-tuned base system, bridging 46\% of the gap between the base system and the oracle system trained with ground truth labels of all data. 

In the rest of this paper, we review the supervised component of our method in Sec.~\ref{s:method}, give detailed description of the proposed method in Sec.~\ref{s:selftrain}, 
compare with related work for semi-supervised ASR in Sec.~\ref{s:related}, 
provide comprehensive experimental results in Sec.~\ref{s:expt}, and conclude with future directions in Sec.~\ref{s:conclusion}.


\section{Supervised learning for ASR}
\label{s:method}

Before describing the proposed method, we 
review the supervised component in our system---CTC with data augmentation.

\subsection{End-to-end ASR with CTC}
\label{s:ctc}
\vspace*{-1ex}

Given an input sequence $X = (\x_1,..., \x_T)$ and the corresponding label sequence $Y = (\y_1,...,\y_L)$, CTC introduces an additional  \texttt{<blank>} token and defines the conditional probability 
\begin{align*}
\bbP (Y|X) = \sum_{p\in \calB^{-1}(Y)}\; \prod_{j=1}^T \bbP (p_j|X)
\end{align*}
where $\calB^{-1}(Y)$ is the set of all paths (frame alignments) that would reduce to $Y$ after removing repetitions and \texttt{<blank>} tokens, and $\bbP (p_j|X)$ is the posterior probability of token $p_j$ at the $j$-th frame by the acoustic model. The underlying assumption is that conditioned on the entire input sequence $X$, the probability for a path $p$ decouples over the frames. 
The CTC loss for one utterance $(X,Y)$ is then defined as $\calL_{CTC} (X, Y) = - \log \bbP (Y|X)$. CTC training minimizes the averaged loss over a set of labeld utterances.
It is well known that after training, the per-frame posteriors from the acoustic model tend to be peaky, and at most frames the most probably token is \texttt{<blank>} with high confidence, indicating ``no emission''.

Due to the abovementioned independence assumption, CTC does not explicitly model transition probabilities between labels, and thus decoding---the problem of $\max_Y \bbP(Y|X)$---is relatively straightforward. The simplest decoder for CTC is the greedy one, which picks the most probably token at each frame and then collapses them by removing repetitions and \texttt{<blank>}'s; we will be mostly using this decoder as it is extremely efficient. One can improve the greedy decoder by maintaining a list of $W$ hypothesis at each frame, leading to a beam search decoder with beamsize $W$. When modeling units are subwords but word-level hypothesis are desired, one can incorporate lexicon and language models, which can be implemented efficiently in the WFST framework~\cite{Miao_15a}. We do not use word-level decoder for generating pseudo-labels since it is much slower than token-level beam search, and it depends on the availability of an in-domain language model. In this work, we only use word-level decoder for evaluating the word error rates (WERs).  
It should be noted that, our self-training method can make use of the attention-based systems~\cite{Chorow_15a,Chan_16a} as well. We use CTC mainly due to its simplicity and efficiency in decoding, for generating pseudo-labels on the fly.

\subsection{Data augmentation}
\label{s:augmentation}
\vspace*{-1ex}


To alleviate the data sparsity issue, a natural approach that does not require unsupervised data is to augment the training data with distorted versions. And various data augmentation techniques have demonstrated consistent improvement for ASR~\cite{Jaitly_13a,Ko_15a,Zhou_17a,Park_19a}. This simple way of obtaining supervised training signal helps us to improve our base system, which in turn generates pseudo-labels with higher quality. 

In this work, we adopt the speed perturbation and spectral masking techniques from~\cite{Park_19a}. Both techniques perturb inputs at the spectrogram level. One can view the input utterance as an image of dimension $D\times T$ where $D$ corresponds to the number of frequency bins, and $T$ the number of frames. Speed perturbation performs linear interpolation along the time axis, as in an image resizing operation; two speed factors $0.9$ and $1.1$ are used. Spectral masking selects $m_F$ segments of the input in the frequency axis with random locations, whose widths are drawn uniformly from $\{0, 1, \dots, n_F\}$, and similarly select $m_T$ segments in the time axis, with widths up to $n_T$. 
We perform grid search of hyperparameters for the supervised CTC system, and set $m_F=1$, $n_F=8$, $m_T=2$, $n_T=16$ throughout. 

\section{Leveraging unsupervised data with self-training}
\label{s:selftrain}
\vspace*{-1ex}

\begin{figure}[t]
    \centering
    \includegraphics[width=1\linewidth,bb=0 30 750 400,clip]{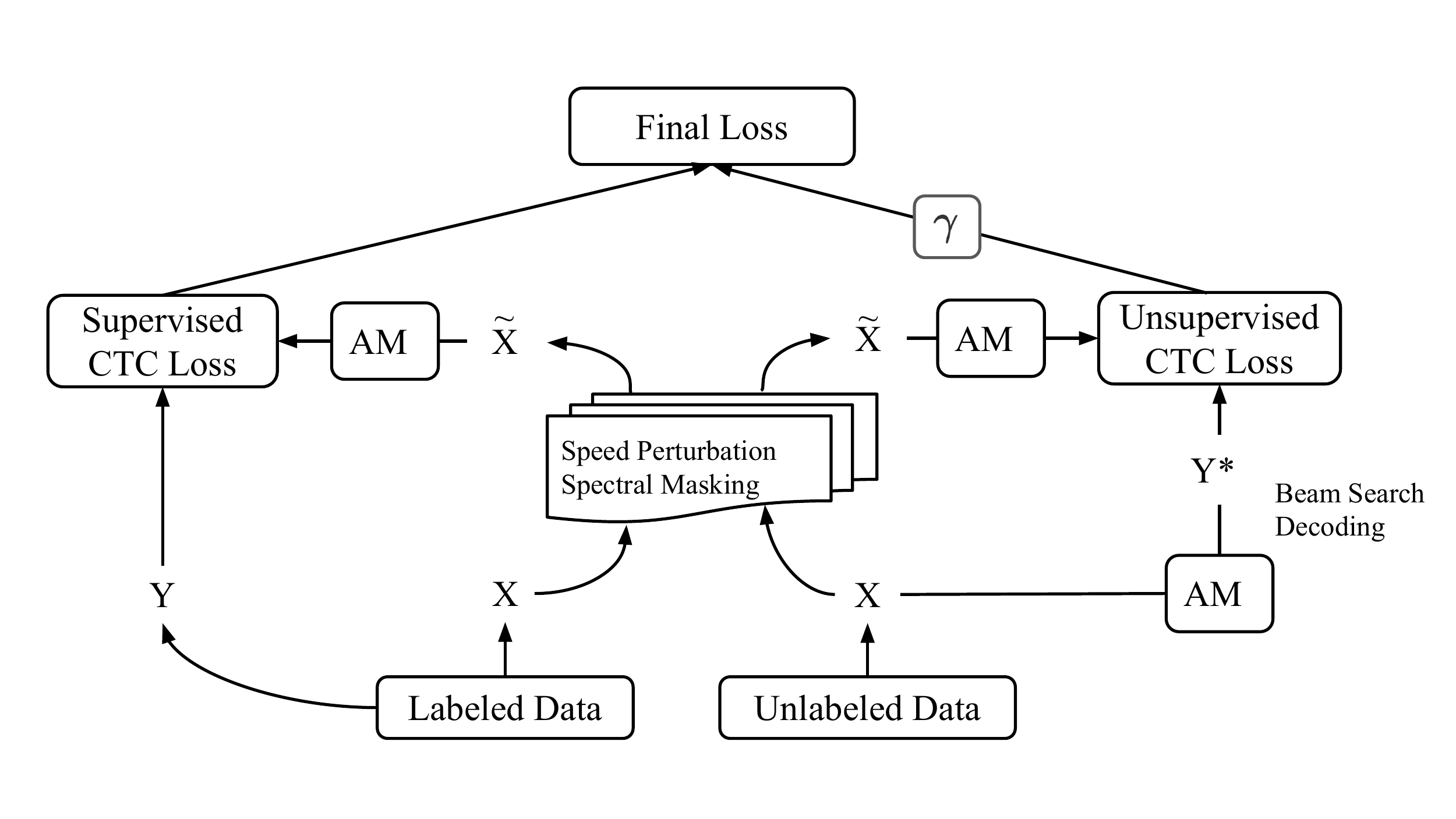}
    \vspace{-2ex}
    \caption{Our self-training method for semi-supervised ASR. }
    \label{f:uda}
    \vspace{-2ex}
\end{figure}

After a base system is sufficiently trained on supervised data, it can be used to predict labels on the originally non-transcribed data. If we take the confident predictions and assume that they are correct, we can add the input and the predictions (pseudo-labels) into training. If the noise in pseudo-labels is sufficiently low, the acoustic model can benefit from the additional training data to obtain improved accuracy. 
We propose to repeat the pseudo-label generation and augmented training steps, so as to have the two reinforce each other, and to continuously improve both. 
In our method, for each update, we generate pseudo-labels for a mini-batch of unsupervised utterances using the current acoustic model with beam search, and compute the CTC losses for these utterances based on their most probable hypothesis. The losses for unsupervised utterances are discounted by a factor $\gamma>0$ to accommodate label noise, and combined with the CTC loss for supervised data to derive the next model update. 
A schematic diagram of our 
method is provided in Fig.~\ref{f:uda}.

Equivalently, we can formulate our method as minimizing the following objective:
\begin{align} \label{e:obj-self}
\min_{\theta, \{Y_j^*\}} \;  \frac{1}{N_l} \sum_{i=1}^{N_l} \calL (X_i, Y_i; \Theta)  +
\frac{\gamma}{N_u} \sum_{j=1}^{N_u} \calL (X_j, Y_j^*; \Theta)
\end{align}
where $\calL(X,Y)$ denotes the CTC loss, we have $N_l$ supervised utterances and $N_u$ unsupervised utterances, $\Theta$ denotes weight parameters in the acoustic model, and we also include the (non-observed) label sequences $\{Y_j^*\}$ of unsupervised utterances as variables. This is a well-defined learning objective, and our method effectively performing alternating optimization over $Y_j^*$ (by beam search) and the weights $\Theta$ (by gradient descent) over mini-batches. 
Additionally, we can perform data augmentation on the unsupervised data, by using the label sequence decoded from the original data on its distorted versions. We will show experimentally that augmenting unsupervised data is as effective as augmenting supervised data.

Our method is motivated by and similar to unsupervised data augmentation (UDA,~\cite{uda}) for semi-supervised learning, in that both methods use pseudo-labels and data augmentation on unsupervised data. But there is a crucial difference between the two: UDA uses soft targets (previous model output) for calculating the unsupervised loss, which encourages the model not to deviate much from that of the previous step, and in fact if there is no data augmentation, the loss on unsupervised data would be zero and has no effect for learning; in contrast, we use the discrete label sequence---output of the beam search decoder on soft targets---on each unsupervised utterance, which provides stronger supervised signals. While~\cite{uda} has not worked on sequence data, we have implemented a sequence version of it, by using the per-frame posterior probabilities as soft targets, and minimizing the cross-entropy loss between soft targets and model outputs at each frame; otherwise the implementation of UDA mirrors that of our method. As demonstrated later, our method outperforms UDA by a large margin.

In view of the peaky per-frame posterior distributions from CTC models,  we think our approach has the advantage that the pseudo-labels are naturally high confidence predictions, relieving us from setting a threshold for discretizing soft probabilities. Although the alignment or locations of non-\texttt{<blank>} tokens can be imprecise from CTC systems, it is not an issue as we only use the label sequence but not its alignment in computing the unsupervised CTC loss, which marginalizes all possible alignments. In this regard, end-to-end systems give a more elegant formulation for self-training, than traditional hybrid systems which rely on alignments.

\section{Related work}
\label{s:related}

Semi-supervised ASR has been studied for a long time, and self-training has been one of the most successful approaches for traditional ASR systems (see, e.g., \cite{long_asr,semi_conf, semi_conf2} and references therein). It is observed that in self-training, the quality of the pseudo-labels plays a crucial role, and much of the research is dedicated to measuring the confidence of pseudo-labels and selecting high confidence ones for supervised training~\cite{semi_conf, semi_conf2}. The issue of label quality becomes even more prominent with LSTM-based acoustic models, which have high memorization capability~\cite{semi_conf3}. In similar spirit, \cite{amz_1m} 
have used a student-teacher learning approach on hybrid systems, to improve accuracy of student using soft targets provided by the teacher on a million hours of non-transcribed data.

Aside from self-training, cycle consistency regularization~\cite{cycle,back_trans} has been applied to semi-supervised ASR. \cite{speech_chain1,speech_chain2, asr_tts1,asr_tts2, asr_tts3} leverage unpaired speech and text data by combining ASR with Text-to-Speech (TTS) modules,
with a training loss that encourages pseudo-labels from ASR to reconstruct audio features well with the TTS system,
and TTS outputs to be recognized by ASR. The authors have proposed different techniques to allow gradient backpropagation through the modules, and to alleviate the audio information loss during text decoding.
Alternatively, \cite{semi_asr} maps audio data with encoder of the ASR model, and maps text with another encoder to a common space, from which text is predicted (from the ASR side) or reconstructed (from the text side) with a shared decoder; 
an additional regularization term is used to encourage representations of paired audio and text to be similar.
The common intuition behind these work is that of auto-encoders, the most straightforward method for unsupervised learning. 
On the other hand, \cite{cri_lm} use adversarial training to encourage ASR output on unsupervised data to have similar distribution as that of unpaired text data, with a criticizing language model.
Our model is much simpler than the above ones, in that we do not have additional neural network models for the text modality; rather, an efficient decoder is used to discretize the acoustic model outputs, and the pseudo-labels are immediately applied to acoustic model training as targets.

Concurrent to our initial submission, the authors of~\cite{fair} also adopted an end-to-end self-training approach. 
A few differences between our work and~\cite{fair} are as follows: first,  we evaluate our method with a CTC-based ASR model whereas they use an attention-based model; second, we use data augmentation on both labeled and unlabeled data and show that both are useful, whereas they do not; third, our method is simpler as we use neither word-level language model nor ensemble methods for generating pseudo-labels; finally, our pseudo-labels are generated on the fly,  where they generate pseudo-labels on the entire unlabeled dataset once. More recent studies~\cite{xu2020,park2020} have similarly shown the effectiveness of 
data augmentation for unsupervised data in self-training.
\section{Experiments}
\label{s:expt}

To demonstrate the effectiveness of the proposed method, we follow a commonly used semi-supervised ASR setup with the WSJ corpus~\cite{asr_tts1,asr_tts2,semi_asr}. We use the \emph{si84} partition (7040 utterances) as the supervised data, and the \emph{si284} partition (37.3K utterances) as unsupervised data. The \emph{dev93} partition (503 utterances) is used as development set for all hyper-parameter tuning, and the \emph{eval92} partition (333 utterances) as the test set.
For input features, we extract 40 dimensional LFBEs with a window size of 25ms and hop size of 10ms from the audio recordings, and perform per-speaker mean normalization. 
We stack every 3 consecutive input frames to reduce input sequence length (after data augmentation), which speeds up training and decoding.

The token set used by our CTC acoustic models are the 351 position-dependent phones together with the \texttt{<blank>} symbol, generated by the Kaldi \emph{s5} recipe~\cite{Povey_11a}. 
Acoustic model training is implemented with Tensorflow~\cite{Abadi_15a}, and we use its beam search algorithm for generating pseudo-labels (with a beamsize $W$) and for evaluating PERs on dev/test (with a fixed beamsize of 20).
To report word error rate (WER) on evaluation sets,  we adopt the
WFST-based framework~\cite{Miao_15a} with the lexicon and the trigram
language model with a 20K vocabulary size provided by the recipe, and
perform beam search using Kaldi's \texttt{decode-faster} with beamsize 20.
Different positional versions of the same phone are merged before word decoding, and we
use the phone counts calculated from \emph{si84} to convert posterior probability (acoustic model output) to likelihood.

Our acoustic model consists of 4 bi-directional LSTM layers~\cite{HochreitSchmid97a} of 512 units in each direction.
For model training, we use ADAM~\cite{KingmaBa15a} with an initial learning rate tuned by grid search. We apply dropout~\cite{Srivas_14a} with rate tuned over $\{$0.0, 0.1, 0.2, 0.5$\}$, which consistently improves accuracy. We use the dev set PER, evaluated at the end of each training epoch, as the criterion for hyperparameter search and model selection. 

\subsection{Base system with data augmentation}
\label{s:base}
\vspace*{-1ex}

As mentioned before, we will use a base system trained only on the supervised data to kick off semi-supervised training. 
For this system, we set the mini-batch size to 4 and each model is trained
up to $40$ epochs. 
We apply data augmentation as described in Sec.~\ref{s:augmentation},
which effectively yields a 3x as large supervised set due to speed
perturbation. In Table~\ref{t:base}, we give PERs of the base system and
another trained without augmentation. Observe that data augmentation provides sizable
gain over training on clean data only (18.52\% vs. 16.83\% for dev PER), leading to higher pseudo-label
quality. We will always use data augmentation on supervised data from now on.

\begin{figure}[t]
\centering
\includegraphics[width=0.8\linewidth,bb=0 10 500 330,clip]{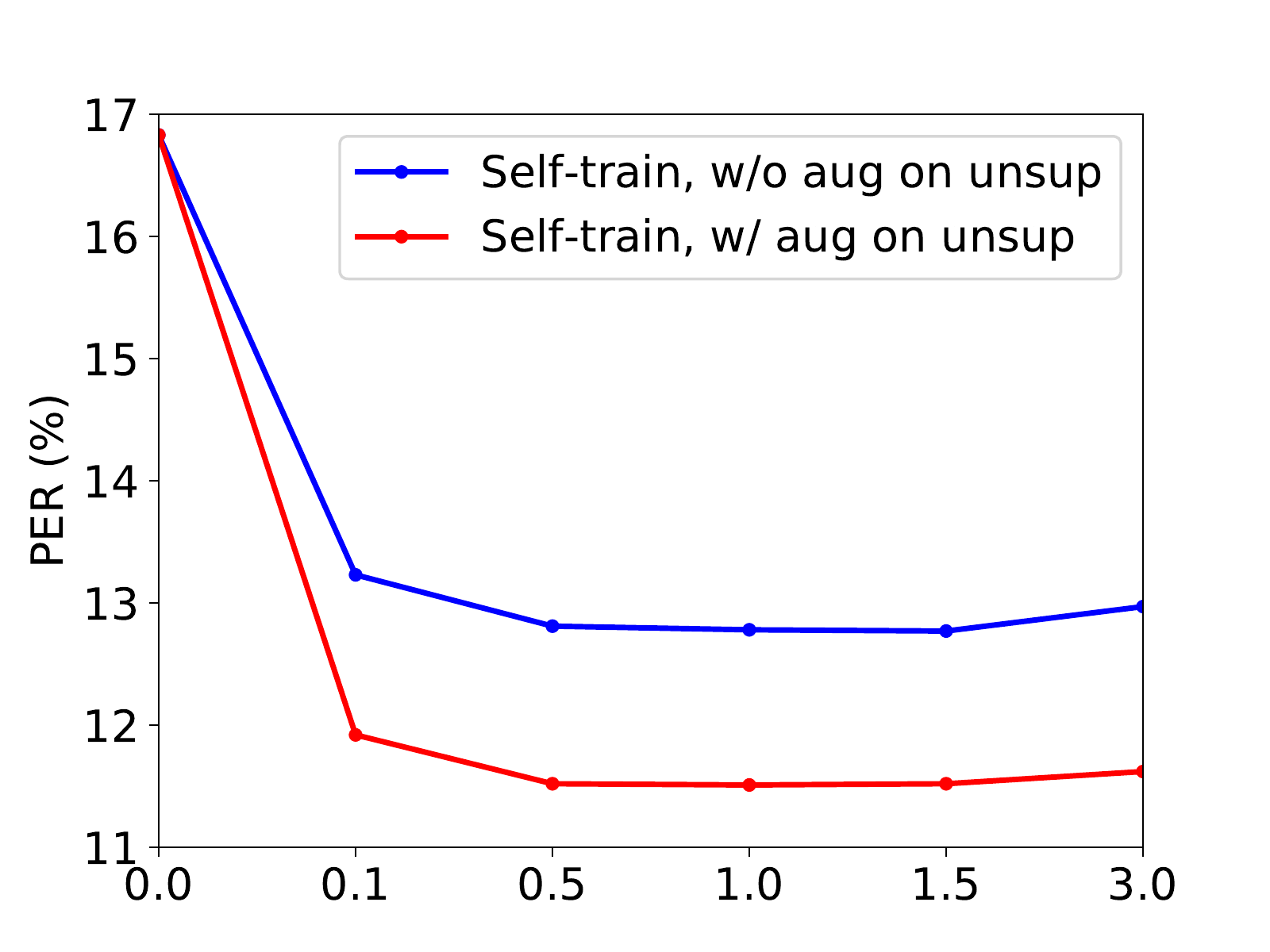}\\
\vspace*{-1ex} $\gamma$\\
\vspace*{-1ex}
\caption{Performance of our method on dev set for different $\gamma$.}
\label{f:vary-gamma}
\vspace*{-2ex}
\end{figure}

\subsection{Continue with self-training}
\label{s:self}
\vspace*{-1ex}

Initialized from the base system, we now continue training with our
semi-supervised objective~\eqref{e:obj-self}. Each model update is
computed with 8 supervised utterances and 32 unsupervised utterances
(since \emph{si284} is about 4 times the size of \emph{si84}, this allows us to process both supervised and unsupervised once in each epoch). The number of unsupervised utterances for each update is not a critical parameter, as the label nosie can be controlled by $\gamma$. 
By grid search, we set the dropout rate to $0.2$, and initial learning rate to $0.0001$
which is 5 times smaller than that for training the initial base model,
and this has the effect of discouraging the model to deviate too much from
the base model. 
Each model is trained for up to another $30$ epochs.
We first set the beam size $W=1$ which corresponds to the greedy decoder,  for generating pseudo-labels on the fly.
We train two set of models, one with data augmentation on unsupervised utterances, and the other one without; but we augment supervised utterances in both cases.
The dev PERs for different values of trade-off parameter $\gamma$ are
given in Fig.~\ref{f:vary-gamma}, and $\gamma=0$ corresponds to the base system.
Our method performs well for a wide range of $\gamma$. The
optimal $\gamma$ is around $1.0$ in both settings, and the performance
does not degrade much with $\gamma>1$, indicating that noise within
pseudo-labels is tolerated to a large degree. 
Furthermore, augmenting the unsupervised data greatly improves the final accuracy.

\begin{table}[t]
\centering
\caption{Performance (measured by \%PER) of different methods on dev and test sets.}
\label{t:base}
\vspace*{-1.5ex}
\begin{tabular}{|l|r|r|}
\hline
Model &\emph{dev93} & \emph{eval92}\\
\hline
CTC w/o DataAug & 18.52 & 13.54 \\
\hline
CTC base system & 16.83 & 11.98 \\
\hline \hline
Self-train W=1 & \textbf{11.51}  & \textbf{8.64} \\
\hspace{5em} w/o DataAug on unsup & 12.77 & \\ 
UDA & 14.27 & \\ 
One-shot pseudo-labels ($W=20$) & 13.68 & \\ 
\hline
\end{tabular}
\vspace*{-1.5ex}
\end{table}

\begin{figure}[t]
\centering
\vspace*{-1ex}
\includegraphics[width=0.9\linewidth,bb=0 0 500 350,clip]{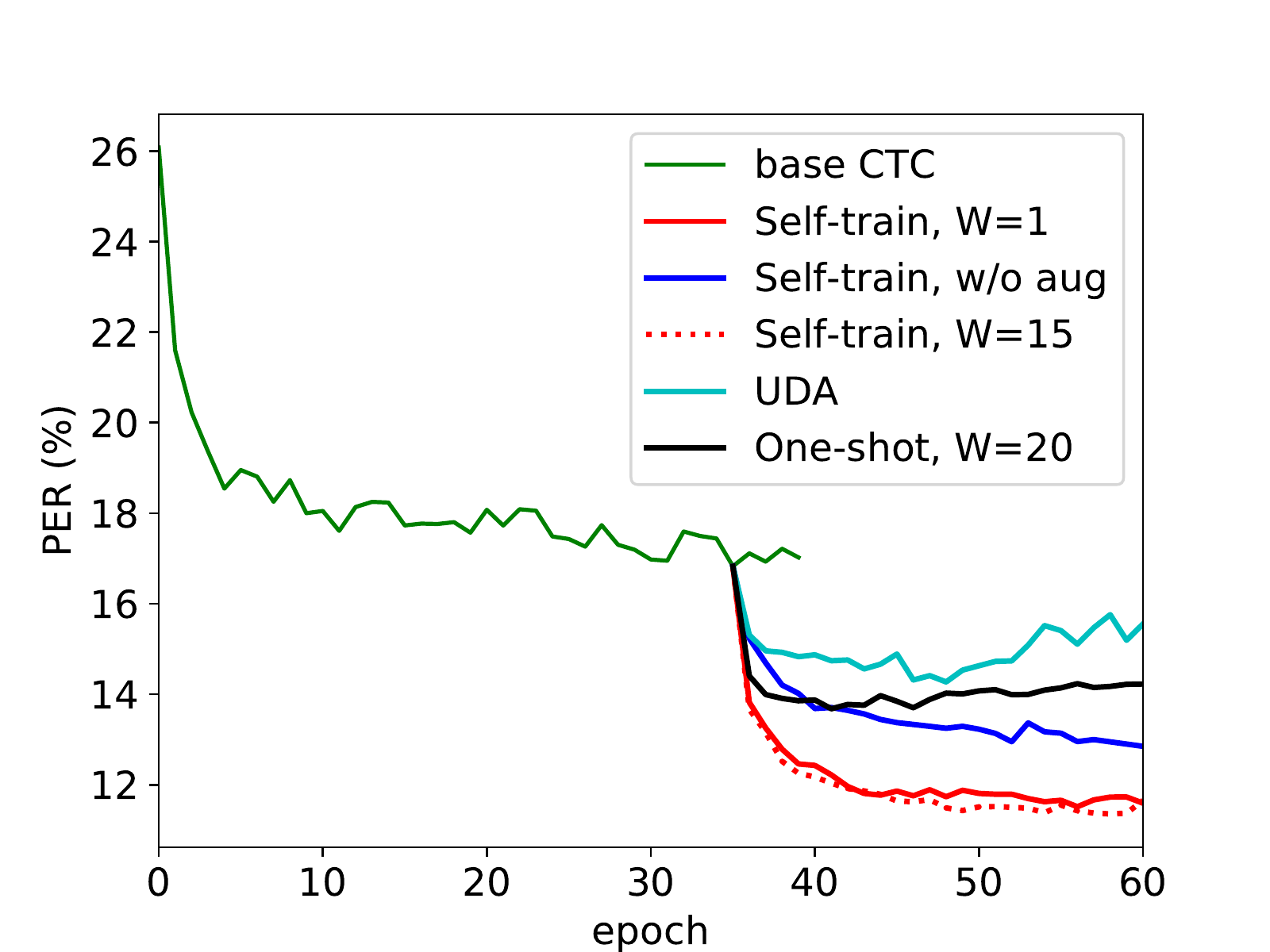}
\vspace*{-1ex}
\caption{Learning curves on dev set for $\gamma=1.0$.
  Semi-supervised learning starts from $36$-th epoch of base model.}
\label{f:learn_curve}
\vspace*{-2ex}
\end{figure}

To show that pseudo-label generation and supervised training
with pseudo-labels reinforces each other, we provide in
Fig.~\ref{f:learn_curve} the learning curve of dev PER vs. epoch for the
models with $\gamma=1.0$. The dev set accuracy improves steadily over
time, with significant PER reductions in the first a few epochs from the base model.

\subsection{Effect of beam size $W$}
\label{s:beamsize}
\vspace*{-1ex}

We now explore the effect of larger $W$, which intuitively shall give higher pseudo-label quality. For this experiment, we fix other hyperparameters to values found at $W=1$. 
In Table~\ref{t:beamsize}, we give the dev PER, as well as the training
time for $W$ in $\{1,\, 5,\, 10,\, 15\}$.
Learning curve with $W=15$ is plotted in Fig.~\ref{f:learn_curve}.
It turns out, with larger $W$, we can slightly improve the final PER, at the cost of much longer training time (mostly from beam search). 
Therefore, we recommend using small $W$ with a good base model.

\begin{table}[t]
\caption{Dev set performance (measured by \%PER) obtained by our method
  with different $W$, together with training times, measured as averaged
  time in seconds spent by each model update, including forwarding and
  decoding the $32$ unsupervised utterances and supervised training on
  $8+32$ utterances. Training time is recorded with a single Tesla K80.}
\label{t:beamsize}
\vspace*{-1.5ex}
\begin{tabular}{@{}|c|c|c|c|c|@{}}
\hline
& $W=1$ & $W=5$ & $W=10$ & $W=15$ \\
\hline
dev PER (\%) & 11.51 & 11.46 & 11.39 & 11.30 \\
\hline
Time / Update &  4.72 & 7.88 & 12.10 & 16.53 \\ 
\hline
\end{tabular}
\end{table}

\subsection{Comparison with UDA}
\label{s:uda}
\vspace*{-1ex}

We now show that hard labels are more useful than soft targets, by
comparing with UDA, which replaces the CTC loss on unsupervised data with
cross-entropy computed with posteriors from previous model. 
We also use data augmentation on unsupervised data, and posteriors are
interpolated in the same way as in speed perturbation for inputs. We tune
the tradeoff parameter $\gamma$ by grid search, and the best performing
model (with $\gamma=0.1$) gives a dev PER of 14.56\% and learning curve in
Fig.~\ref{f:learn_curve}. 
The observation that hard labels outperform soft targets is in line with that of~\cite{guide_ctc2} for teacher-student learning with CTC.

\subsection{Comparison with one-shot pseudo-labels} 
\label{s:one-shot}
\vspace*{-1ex}

To further demonstrate the importance of fresh pseudo-labels, we compare
with a more widely used approach where the pseudo-labels are generated once
on the entire unsupervised dataset with the base model. We do so with a large
decoding beam size $W=20$, and then continue training from the base model
with objective~\eqref{e:obj-self} without updating the pseudo-labels again.
This approach does clearly improve over the base system with a dev PER of
13.68\%, but not as much as our method with $W=1$. Its learning curve is shown in Fig.~\ref{f:learn_curve}, and
the curve plateaus more quickly than those of our method.

\subsection{Results summary}
\label{s:summary}
\vspace*{-1ex}

\begin{table}[t]
\centering
\caption{Performance (measured by \%WER) of previous work and our
  methods on \emph{eval92}. }
\label{t:result_wer}
\vspace*{-1ex}
\begin{tabular}{|l|r|}
\hline
Model & WER \\
\hline
\cite{Baskar_19a} (attention, train on \emph{si84}, & \\ 
$\qquad\qquad\qquad$ unsup on \emph{si284} by ASR+TTS) & 20.30 \\
\hline
RNN-CTC~\cite{GravesJaitly14a}, train on \emph{si84} &  13.50 \\
Our CTC, train on \emph{si84},  w/o DataAug & 13.22  \\
\hline
Base system & 11.43 \\
UDA & 10.93 \\
One-shot pseudo-labels & 10.67 \\
Self-training, $W=1$ &  \textbf{9.78} \\
\hline \hline
EESEN CTC~\cite{Miao_15a}, train on \emph{si284} & 7.87 \\
\hline
\end{tabular}
\vspace*{-1.5ex}
\end{table}

In Table~\ref{t:result_wer} we give WERs of different methods on
\emph{eval92}.
The recent work~\cite{Baskar_19a} which uses the same data partition for
semi-supervised learning with attention models is also included.
To put our results in close context, we have included the CTC model
from~\cite{GravesJaitly14a} trained on \emph{si84} only. 
Our method with $W=1$ gives a relative 31.6\% dev PER reduction
(16.83\%$\rightarrow$11.51\%), and a relative 14.4\%
test WER reduction (11.43\%$\rightarrow$9.78\%) over a carefully-trained
base system with data augmentation, effectively reducing the performance
gap between the base system (11.43\%) and the oracle system (7.87\%) by 46\%.

\section{Future directions}
\label{s:conclusion}

As for future directions, we believe that word-level decoding, which incorporates lexicon and an in-domain language model, can further improve the quality of pseudo-labels after converting the word sequence back to token sequence (see, e.g.,~\cite{data_tech}), at the cost of longer decoding time. Another promising model to be used in our method is RNN-transducer~\cite{Graves_12a}, which has a built-in RNN LM to model label dependency and to improve token-level decoding.
Furthermore, for larger $W$ one may consider the top a few hypotheses, and use all of them for computing the loss on unsupervised data~\cite{distill, fair}.


\bibliographystyle{IEEEtran}
\bibliography{refs}
\end{document}